# Skyrmion-Bubble Bundles in an X-type $Sr_2Co_2Fe_{28}O_{46}$ Hexaferrite above Room Temperature

*Jin Tang, Yaodong Wu\*, Jialiang Jiang, Lingyao Kong, Shouguo Wang, Mingliang Tian, and Haifeng Du\**

Jin Tang, Lingyao Kong, Mingliang Tian

School of Physics and Optoelectronics Engineering, Anhui University, Hefei, 230601, China

Yaodong Wu

School of Physics and Materials Engineering, Hefei Normal University, Hefei, 230601, China

E-mail: wuyaodong@hfnu.edu.cn

Jin Tang, Jialiang Jiang, Mingliang Tian, Haifeng Du

Anhui Province Key Laboratory of Condensed Matter Physics at Extreme Conditions, High Magnetic Field Laboratory, HFIPS, Anhui, Chinese Academy of Sciences, Hefei, 230031, China

E-mail: duhf@hmfl.ac.cn

Shouguo Wang

School of Materials Science and Engineering, Anhui University, Hefei 230601, China

## Abstract

Magnetic skyrmions are spin swirls that possess topological nontriviality and are considered particle-like entities. They are distinguished by an integer topological charge $Q$. The presence of skyrmion bundles provides an opportunity to explore the range of values for $Q$, which is crucial for the advancement of topological spintronic devices with multi-$Q$ properties. In this study, we present a new material candidate, $Sr_2Co_2Fe_{28}O_{46}$ hexaferrite of the X-type, which hosts small dipolar skyrmions at room temperature and above. By exploiting reversed magnetic fields from metastable skyrmion bubbles at zero fields, we can incorporate skyrmion-bubble bundles with different interior skyrmion/bubble numbers, topological charges, and morphologies at room temperature. Our experimental findings are consistently supported by micromagnetic simulations. Our results highlight the versatility of topological spin textures in centrosymmetric uniaxial magnets, thereby paving the way for the development of room-temperature topological spintronic devices with multi-$Q$ characteristics.

## 1. Introduction

Magnetic skyrmions are spin swirls that act like particles and are characterized by a topological charge $Q$, which is a quantized integer that remains constant even during continuous deformation of the spin configuration[1, 2]. Skyrmions have great potential as carriers of information for use in memory and processing devices due to their unique topology-related electromagnetic properties[3-9]. Recent studies have revealed that skyrmion bundles -also called skyrmion bags-, which consist of numerous skyrmions encircled by a closed spin spiral, can extend the range of topological charges from 1 to any integer values[10-13]. Skyrmion bundles have been first observed in the chiral magnet FeGe at 95 K[10], and this observation has opened up opportunities for developing topological spintronic devices such as multi-bit memory and information multiplexing based on the freedom parameter of $Q$.[11, 14, 15] Applying a reversed magnetic field in the mixed zero-field skyrmion-helix state can control and create skyrmion bundles, and skyrmion bundles have also been then observed in FeGeTe below 200 K[16].

However, practical spintronic devices typically operate at room temperature or higher, which means that materials that can host skyrmions with high curie temperature $T_c$ are needed[6, 17, 18]. A recent study tested the use of reversed magnetic fields to create skyrmion bundles in the room-temperature skyrmion material CoZnMn alloys, but only the melting of the skyrmion lattice was observed, with no skyrmion bundles appearing[19]. Dipolar skyrmions - also known as type-I bubbles or skyrmion bubbles - in centrosymmetric uniaxial magnets are topologically equivalent to chiral skyrmions[20-30]. Dipolar skyrmions share the same topology[29, 30], similar topological magnetism[22, 25], and comparable nanometric size at room temperature as chiral skyrmions[21, 26]. Moreover, there is a wide range of uniaxial magnets that can host dipolar skyrmions at room temperature and above[21, 24-27, 29], which raises the possibility of exploring skyrmion bundles in these uniaxial dipolar magnets at room temperature.

In this study, we report the discovery of skyrmion-bubble bundles in a centrosymmetric uniaxial magnet known as the X-type $Sr_2Co_2Fe_{28}O_{46}$ (SrCoFeO) hexaferrite[31-34]. SrCoFeO hexaferrite is a new material candidate that hosts dipolar skyrmions with nanometric-scale sizes (~150-250 nm) at room temperature. By using reversed magnetic fields, we can create skyrmion-bubble bundles in which various internal skyrmions and bubbles are surrounded by an outer boundary stripe domain. The skyrmion-bubble bundles found in SrCoFeO exhibit distinct characteristics compared to skyrmion bundles found in chiral magnets[10]. These include exceptional stability even at room temperature, a wide range of styles in terms of the arrangement of skyrmions and bubbles, and the ability to manipulate the topological properties through the application of external tilted fields.

## 2. Results and Discussions

### 2.1. Magnetic properties of the bulk SrCoFeO single crystal

Hexaferrites are a significant material system with a super-high Curie temperature $T_c$ and a wide range of applications, including hard magnets and multiferroic materials[31-40]. The abundant anisotropy behavior of hexaferrites predicts that they have the potential to host dipolar skyrmions at room temperature. To date, the M-type BaFeScO magnet is the only hexaferrite system known to host dipolar skyrmions[35, 36]. Here, we demonstrate the stabilization of topological spin textures in the X-type SrCoFeO hexaferrite. We prepared bulk SrCoFeO single crystals using the standard flux method[33, 34].

**Figure 1**a shows the crystal model of the X-type SrCoFeO hexaferrite[33, 34], which possesses a crystal structure belonging to the trigonal system, with a space group of $R\bar{3}m$. The crystal constants for this structure are reported as $a = 0.59$ nm and $c = 8.41$ nm. The $c$-cut SrCoFeO crystals exhibit good crystallinity, as revealed by X-ray diffraction, electronic diffraction, and high-resolution transmission electron microscopy (TEM) (Supplemental Figure S1). To analyze the macroscopic magnetic properties of the synthesized SrCoFeO crystals at room temperature and above, we employed the Quantum Design Physical

Properties Measurement System (PPMS), as shown in Figure 1. During the magnetic field ($B$) cooling process at $B = 10$ mT (Figure 1b), the magnetization indicates a Curie temperature of approximately 750 K. The magnetizations as a function of the magnetic field (Figure 1c and Supplemental Figure S2) suggest that the magnetization easy axis is along the $c$-axis at room temperature and above, which aligns with previous studies[31, 34]. By applying a magnetic field ranging from zero to the saturation field $B_{sat}$, the total energy provided by the magnetic field can be expressed as $E_B = \int_0^{B_{Sat}} M(B) dB$. This energy must surpass both the magnetocrystalline anisotropy energy and the demagnetization energy. The demagnetization energy is primarily influenced by the sample's shape and results from magnetic dipole-dipole interactions. In our macroscopic measurements of magnetizations, we utilize a cube-shaped sample, resulting in comparable demagnetization energies along and perpendicular to the easy magnetization $c$-axis[41]. To determine the uniaxial magnetic anisotropy ($K_u$), we calculated the differential values of $\int_0^{B_{Sat}} M(B) dB$ between along and perpendicular to the $c$ axis, as marked by the shaded area in Figure 1c[42]. Both the uniaxial magnetic anisotropy and effective anisotropy field ($H_k = \frac{2K_u}{u_0 M_s}$) initially increase with temperature and then decrease near $T_c$ (Figure 1d). The saturation magnetization ($M_s$) is taken as the magnetization at $B = 1$ T along the $c$-axis. The quality factor ($\eta = \frac{2K_u}{u_0 M_s{}^2}$), which characterizes the stability of magnetic bubbles in uniaxial magnets, increasing from approximately 0.8 at 300 K to 9 at 725 K (Figure 1e).

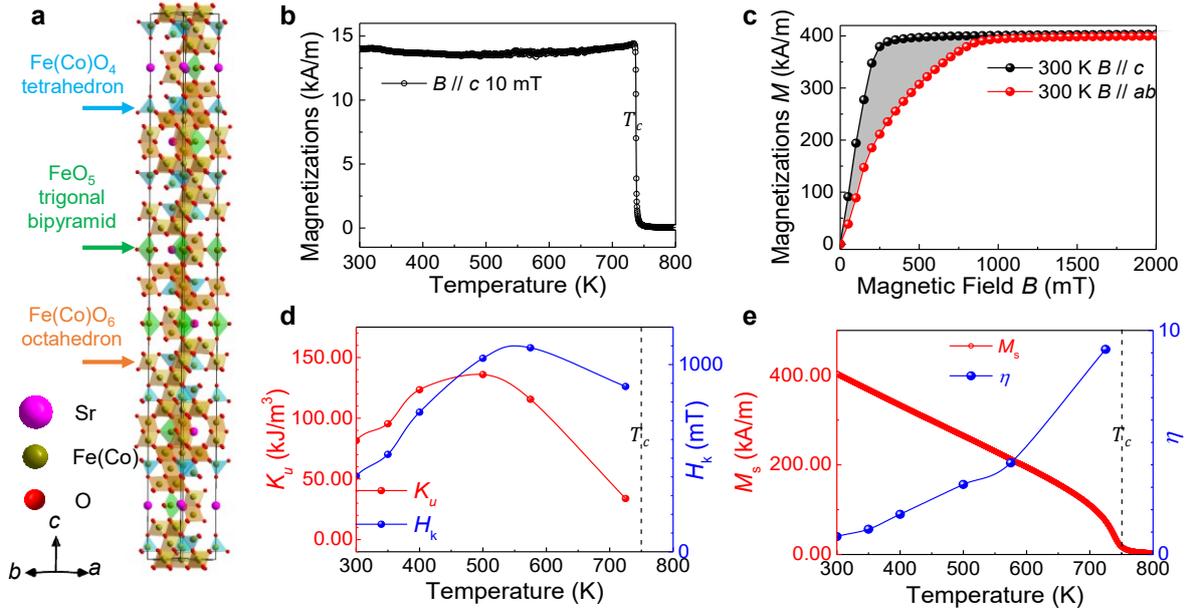

**Figure 1** a) Crystal structure model of the X-type SrCoFeO hexaferrite. b) Magnetization $M$ in the field cooling process at $B = 10$ mT applied along the $c$ axis. c) Magnetization as a function of the magnetic field at $T = 300$ K. d) Magneti anisotropy $K_u$ and anisotropy field $H_k$ as a function of temperature $T$. e) Saturation magnetization $M_s$ and quality factor $\eta$ as a function of temperature $T$. $M_s$ is taken as the magnetization in the field cooling process at $B = 1$ T applied along the $c$ axis.

## 2.2. Observation of dipolar skyrmions in SrCoFeO lamella

Firstly, we examined the magnetic domains in $c$-cut SrCoFeO lamella using Lorentz-TEM, which allows visualization of the in-plane magnetizations ($m_{xy}$)[43]. In the as-fabricated 150-nm thick SrCoFeO lamella at zero fields and 294 K (**Figure 2**a), typical stripe domains emerged, stabilized by the interplay between perpendicular magnetocrystalline anisotropy and dipole-dipole interactions. As an out-of-plane magnetic field is applied, the stripe domains gradually shrink and transform to dipolar skyrmions and bubbles at high fields. Due to the absence of chiral Dzyaloshinskii-Moriya (DM) interaction in the centrosymmetric hexaferrite, coexisting skyrmions with counterclockwise (black under-defocus Fresnel contrasts) and clockwise (white under-defocus Fresnel contrasts) rotations are observed (Figure 1b)[24]. By employing measured magnetic parameters in simulations, simulated dipolar skyrmions and bubbles are found to exhibit comparable sizes and magnetic contrasts (Figure 2c and Supplemental Figure S3) to the experimental observations. Skyrmions display closed domain walls, whereas bubbles featured a pair of Bloch lines within their domain walls (Supplemental Figure S3)[22, 24]. The size of the skyrmions, ranging from ~150-250 nm, is comparable to that observed in chiral CoZnMn (~115-187 nm)[6], Mn-Pt-Pd-Sn (~>135 nm)[44], FeNiPdP (~>200 nm)[45], and magnetic multilayer films at room temperature (Figure 2d)[17]. By varying the magnetic field at different temperatures around room temperature starting from the initial stripe domains, we verified the robust stabilization of dipolar skyrmions near room temperature. Notably, at temperatures below approximately 260 K, only topologically-trivial bubbles are observed in the SrCoFeO magnet due to the temperature-induced spin reorientation effect. This reorientation causes the magnetization easy axis to tilt away from the $c$-axis[34]. A previous study has demonstrated the transformation of skyrmions to bubbles when the easy axis is tilted away from the out-of-plane direction[46]. The topologically trivial bubbles become the more stable phases because the magnetization along the easy uniaxial axis is favored by the domain wall magnetizations of the cylinder domains. This preference helps minimize the energy associated with the uniaxial magnetic anisotropy[46]. Our simulation also reveals the stabilization of bubbles in the presence of a titled uniaxial magnetic anisotropy (see Supplemental Figure S4).

We also investigate the magnetic domains in a 150-nm thick lamella where the $c$-axis is parallel to the plane, as shown in Supplemental Figures S5 and S6. At temperatures above 260 K, we observe the presence of multi-domains with magnetizations aligned along the two axes of the easy $c$-axis. By applying out-of-plane magnetic fields, we observe the displacement of domain walls, leading to a single domain state at high-field strengths. In this case, we do not observe any skyrmions or bubbles due to the absence of perpendicular anisotropies. Nevertheless, micromagnetic simulations successfully replicate the formation and magnetic evolution of these multi-domains under various field strengths. Furthermore, we can also identify the temperature-induced spin reorientation effect in the lamella, where the $c$-axis is parallel to the plane. At low temperatures, the magnetizations of the multi-domains deviate from alignment with the *c*-axis. These magnetic domain evolutions of the lamella with the *c*-axis parallel to the plane are consistent with a previous study[34].

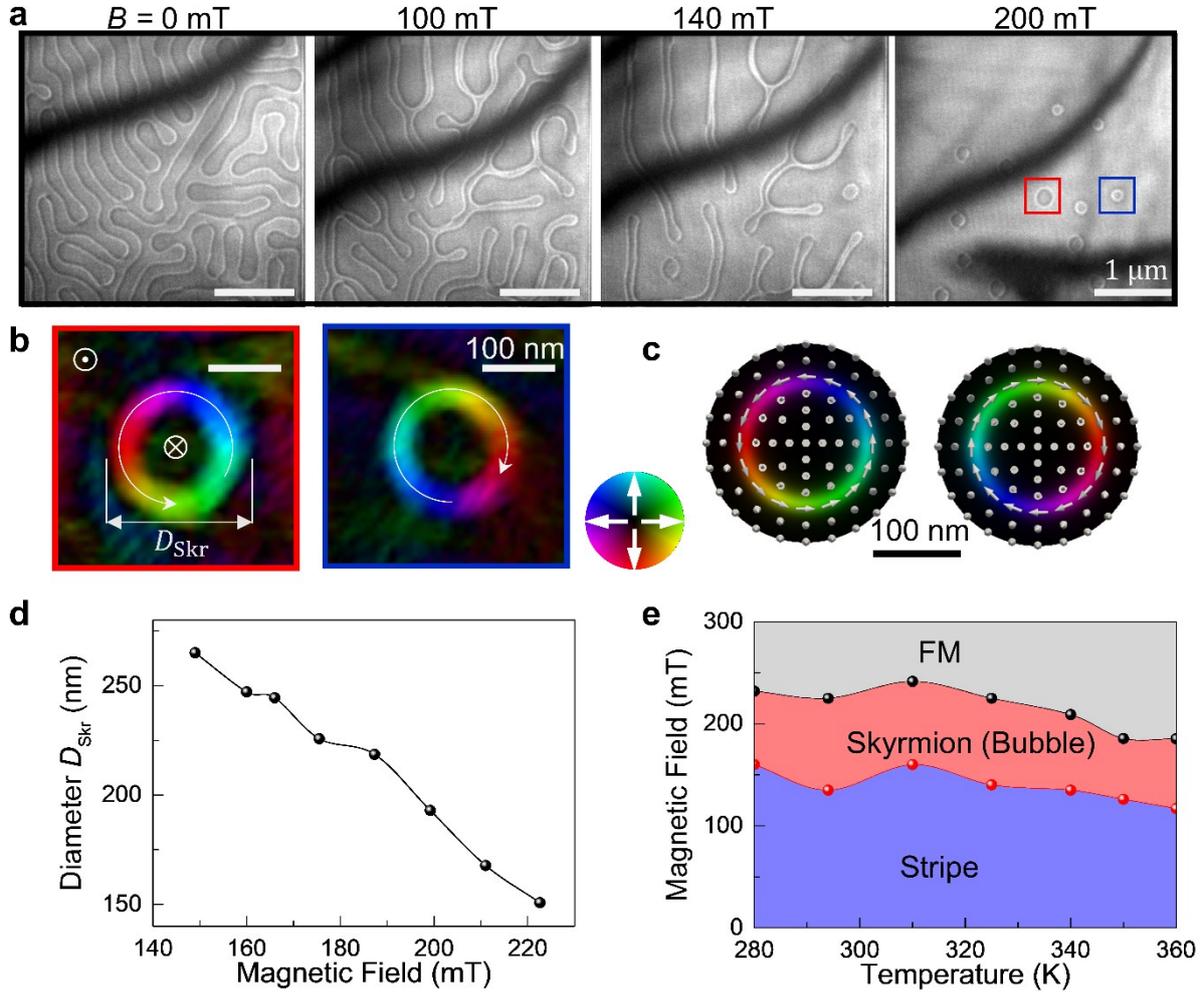

**Figure 2** Observation of dipolar skyrmions above room temperature. a) Magnetic evolution from zero-field stripe domains driven by out-of-plane magnetic fields at $T = 294$ K. Scale bar, 1 μm. b) Representative in-plane magnetization mapping of dipolar skyrmions retrieved from TIE analysis. Scale bar, 100 nm. c) Simulated in-plane spin configurations for the dipolar skyrmions based on measured magnetic parameters at room temperature. d) Diameter of the skyrmion $D_{Skr}$ as a function of the magnetic field at $T = 294$ K. e) Magnetic phase diagram of magnetic domains as a function of temperature and magnetic field. The red and black dots represent the threshold magnetic fields for stripe to skyrmion and skyrmion to FM transformations, respectively, in the field-increasing process. The color represents the in-plane magnetization according to the colorwheel.

## 2.3 Observation of skyrmion-bubble bundles in SrCoFeO lamella

Although stripe/helix domains are known to be the spontaneous magnetic ground states at zero magnetic fields, skyrmions or stripy-skyrmions can also exist as metastable phases at zero fields[10, 47, 48]. By applying reversed magnetic fields, $Q = 1$ skyrmions can persist and be surrounded by a closed spiral, leading to the formation of skyrmion bundles[10]. The observation of skyrmion bundles has opened up possibilities for creating intriguing topological spin textures, including skyrmion-antiskyrmion pairs and Hopfions, by manipulating reversed fields applied to zero-field skyrmions[49, 50]. Due to the intricate coexistence of skyrmions and bubbles in the centrosymmetric SrCoFeO magnet, we define the skyrmion bundles as the bundles wherein both the outer boundary and internal constituents are exclusively composed of skyrmions. Similarly, bubble bundles are defined as bundles wherein both the outer boundary and internal constituents solely consist of bubbles. Furthermore, the skyrmion-bubble bundle is defined as the bundle that encompasses main constituents containing both skyrmions and bubbles.

To achieve these bundles, we first applied a negative field to create skyrmions and bubbles. Subsequently, we decreased the field to 0 mT, resulting in the formation of mixed stripes and stripy-skyrmions[47]. Finally, by applying positive magnetic fields, we can assemble various skyrmions and bubbles encircled by a larger closed stripe domain (**Figure 3**a and Supplemental Figure S7), referred to as skyrmion-bubble bundles in this study. Our simulations (Figure 3b and Supplemental Figure S8) successfully reproduce the formation of skyrmion-bubble bundles observed in experiments, validating the reliability of using reversed magnetic fields to create skyrmion-bubble bundles.

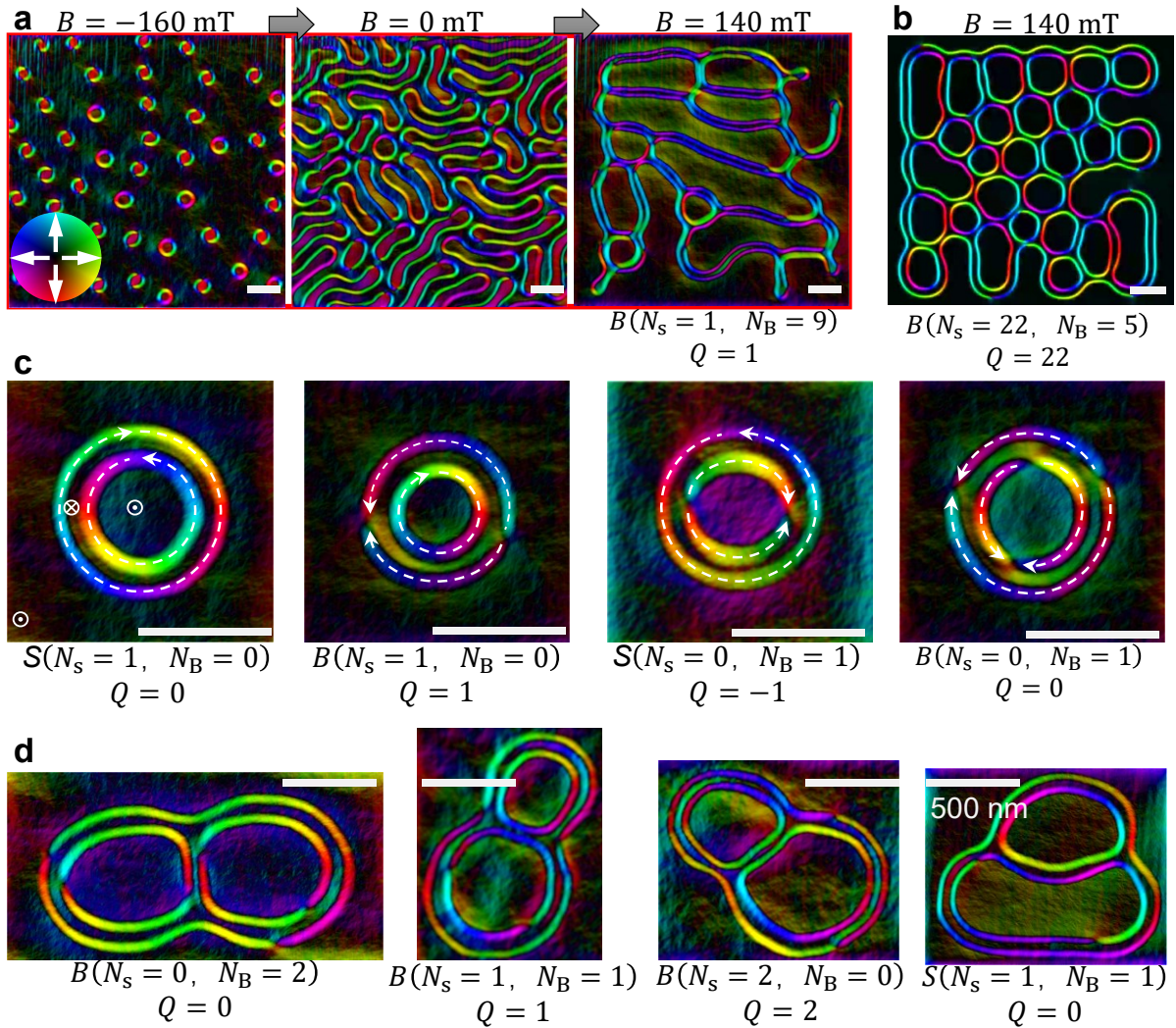

**Figure 3** Observation of skyrmion-bubble bundles. a) Skyrmion-bubble bundles obtained by reversed magnetic fields. b) Simulated skyrmion-bubble bundles based on measured magnetic parameters at 350 K. c) Diversity of skyrmion-bubble bundles with one interior bubble, which are known as $2\pi$-vortices. d) Diversity of skyrmion-bubble bundles with two interior bubbles. Temperature $T = 350$ K. Scale bar, 500 nm. The color represents the in-plane magnetization according to the colorwheel.

Skyrmion-bubble bundles, in contrast to chiral skyrmion bundles, are stabilized by the magnetic dipole-dipole interaction and offer additional degrees of freedom in terms of the distribution of skyrmions and bubbles. Let's consider the simple skyrmion-bubble bundles with 1 internal skyrmion/bubble, known as the $2\pi$-vortex[51, 52], as an example. There are four different styles: a skyrmion encircled by a skyrmion (also called skyrmionium) with $Q = 0$[53, 54], a skyrmion encircled by a bubble with $Q = 1$, a bubble encircled by a skyrmion with $Q = -1$, and a bubble encircled by a bubble with $Q = 0$ (Figure 3b and Supplemental Figure S9).

The diversity of $2\pi$-vortices, despite having the same size but different topologies and configurations, makes them ideal candidates for applications in multi-state information processing and memory. In contrast, there is only one style $2\pi$-vortex in chiral magnets[10].

Here, we use the notation $S(N_S,\ N_B)$ and $B(N_S,\ N_B)$ to represent the skyrmion-bubble bundles. The outer boundary stripe domain of $S(N_S,\ N_B)$ does not contain Bloch lines, while that of $B(N_S,\ N_B)$ includes Bloch lines. The emergence of Bloch lines contributes $Q = 0$ as that of the magnetic bubbles (Supplemental Figure S3). The counts of internal skyrmions and bubbles are represented by $N_S$ and $N_B$, respectively, and the corresponding topological charges are given by $Q = N_S - 1$ and $Q = N_S$ for $S(N_S,\ N_B)$ and $B(N_S,\ N_B)$ skyrmion-bubble bundles, respectively. Considering a skyrmion-bubble bundle consisting of $N_S$ skyrmions and $N_B$ bubbles, with $N = N_S + N_B$, there exist a total of $2(N + 1)$ distributions of skyrmions and bubbles with topological charges $Q$ varying from $-1$ to $N$. Taking the example of skyrmion-bubble bundles with $N = 2$, we can observe 6 distinct types of bundles: $S(N_S = 2, N_B = 0)$ with $Q = 1$, $S(N_S = 1, N_B = 1)$ with $Q = 0$, $S(N_S = 0, N_B = 2)$ with $Q = -1$, $B(N_S = 2, N_B = 0)$ with $Q = 2$, $B(N_S = 1, N_B = 1)$ with $Q = 1$, and $B(N_S = 0, N_B = 2)$ with $Q = 0$. Our experiments involved the repetition of the process of reversing fields, which enable us to obtain skyrmion-bubble bundles exhibiting a wide range of diversity (See Figure 3 and Supplemental Figure S10).

**2.4 Topological magnetic transformations of skyrmion-bubble bundles**

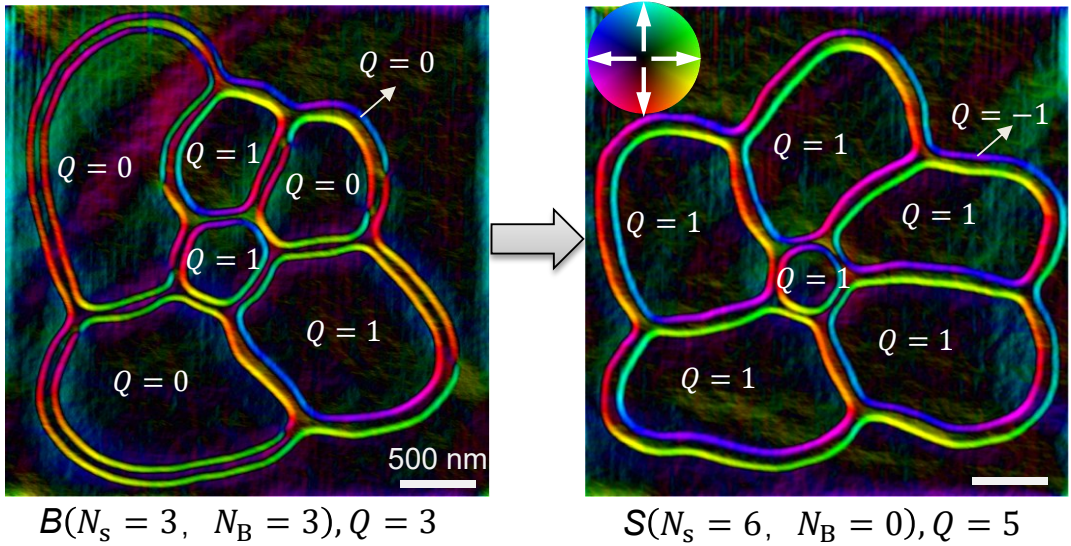

$B(N_S = 3, \ N_B = 3), Q = 3$   $S(N_S = 6, \ N_B = 0), Q = 5$

**Figure 4** Topological transformation from skyrmion-bubble bundles to skyrmion bundles with an increased $Q$ by a wobbler operation at a fixed $B = 140$ mT. Temperature $T = 350$ K. The color represents the in-plane magnetization according to the colorwheel.

Since coexisting skyrmions and bubbles are typically obtained under negative fields, we usually obtain skyrmion-bubble bundles with $Q = N_S$ when reversing the magnetic fields, denoted by $B(N_S, \ N_B)$. The Bloch lines, which lie within the domain walls of bubbles, move with titled in-plane magnetic fields (Supplemental Figure S11). During their rapid motion induced by a wobbler operation, wherein the tilted angle is increased gradually from 0° to 15°, then to $-15°$ and finally back to 0°, Bloch lines become unstable and can annihilate (Supplemental Figure S8)[25, 55]. Therefore, the wobbler operation can be considered a viable method for achieving topological magnetic transformation from a skyrmion-bubble bundle $B(N_S, \ N_B)$ to a dipolar skyrmion bundle $S(N_S = N, \ N_B = 0)$. **Figure 4** shows an example of such a transformation, depicting the conversion of a $B(N_S = 3, \ N_B = 3)$ skyrmion-bubble bundle to a $S(N_S = 6, \ N_B = 0)$ skyrmion bundle. Similar to chiral skyrmion bundles, the topological charge of a dipolar skyrmion bundle $S(N_S = N, \ N_B = 0)$ is equal to $Q = N - 1$, where $N$ represents the internal skyrmion number (Supplemental Figure S12). The outer closure stripe domain contributes to the negative integer charge. In principle, the internal count of skyrmions can be any integer $N \geq 1$. Thus, $Q$ of skyrmion bundles can be any integer $Q \geq 0$ at positive magnetic fields. It should be noted that the magnetizations between the internal skyrmion and the outer boundary have orientations opposite to that of the magnetic field. Therefore, the internal skyrmions/bubbles of bundles always maintain tight contact to minimize the Zeeman energy, even when initially arranged in a loosely packed configuration (Supplemental Figure S12).

Once the high-$Q$ skyrmion bundles stabilize, increasing the magnetic fields will reduce and eventually eliminate skyrmions, leading to a decrease in topological charges $Q$, as shown in **Figure 5**. For instance, a bundle with $Q = 6$ converts to a bundle with $Q = 3$ at $B = 151$ mT, $Q = 2$ at $B = 155$ mT, $Q = 1$ at $B = 156$ mT and $Q = 0$ (*i.e.* skyrmionium) at $B = 162$ mT. Finally, at $B = 164$ mT, a skyrmion with $Q = -1$ is obtained.

We thus can apply the wobbler (Figure 4) and $B$-increasing (Figure 5) operations to obtain high-$Q$ and low-$Q$ skyrmion bundles, respectively. By combining these operations with reversed magnetic fields, we can obtain dipolar skyrmion bundles with various topological charges $Q$, as shown in **Figure 6** and Supplemental Figure S13. Moreover, it is worth noting that skyrmion bundles can also stabilize at zero magnetic fields, albeit with distorted shapes (Supplemental Figure S14). As an example, skyrmionium can stabilize across a broad field-temperature range, and at high fields, the skyrmionium transforms to a skyrmion (Figure 6c and Supplemental Figure S14).

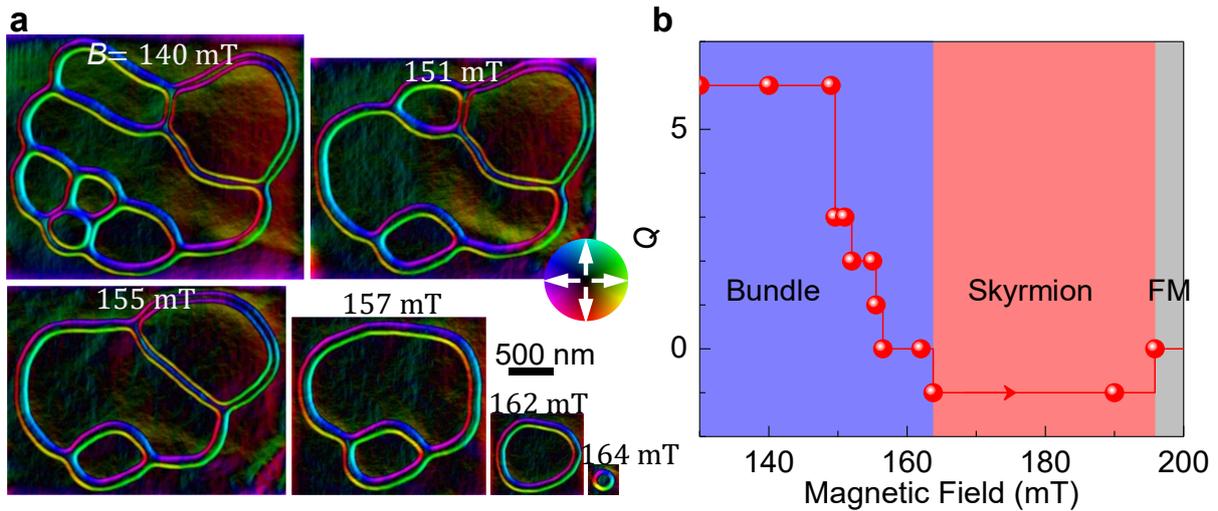

**Figure 5** Decrease of interior skyrmion number and topological charge $Q$ in the field-increasing operation. (a) Magnetic evolution from a $Q = 6$ skyrmion bundle driving by an out-of-plane magnetic field. (b) Topological charge $Q$ as a function of the magnetic field. Scale bar, 500 nm. Temperature $T = 350$ K. The color represents the in-plane magnetization according to the colorwheel.

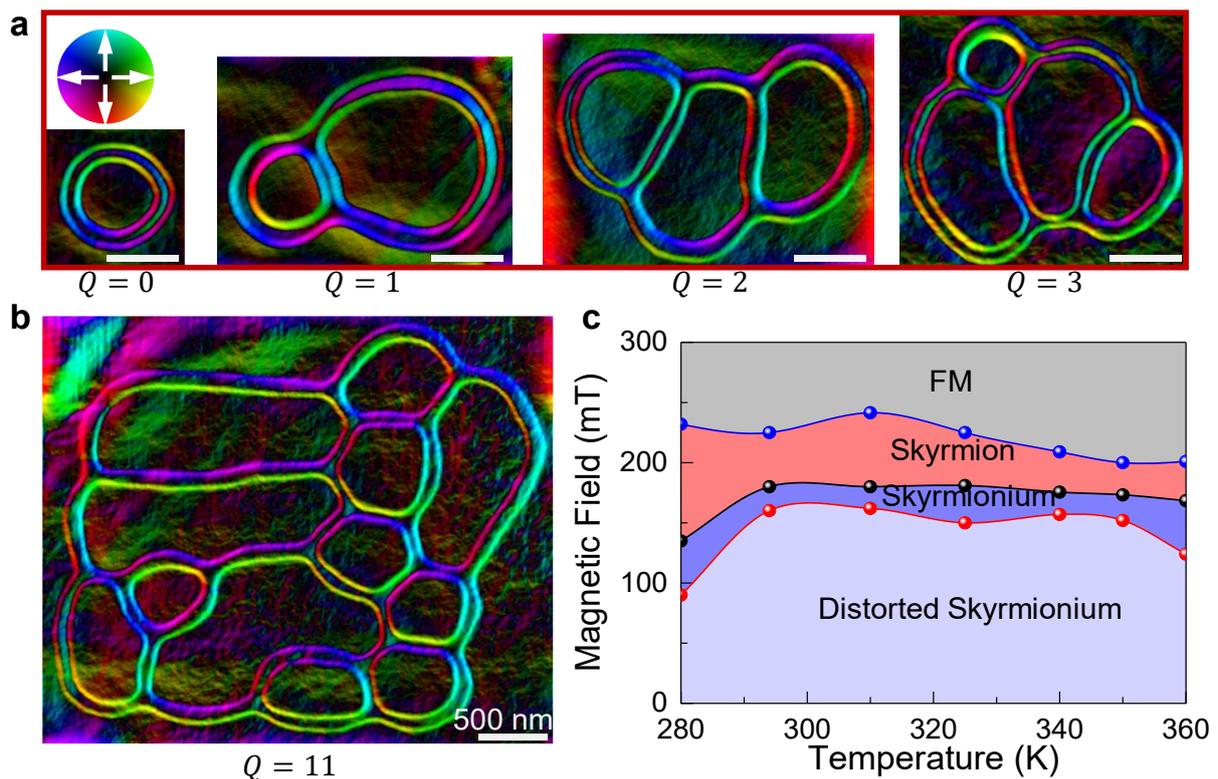

**Figure 6** Diversity of topological charges $Q$ for dipolar skyrmion bundles. $B = 140$ mT. a) Skyrmion bundles with $Q = 0, 1, 2, 3$. b) Skyrmion bundles with $Q = 11$. c) Magnetic phase diagram for the $Q = 0$ bundle, *i.e.* the skyrmionium, as a function of magnetic field and temperature. Scale bar, 500 nm. The color represents the in-plane magnetization according to the colorwheel.

## 3. Conclusions

In summary, we have presented a new material - the X-type SrCoFeO hexaferrite - that hosts nanometric-scale (~150-250 nm) dipolar skyrmions and topological skyrmion-bubble bundles at room temperature and higher. By applying reversed magnetic fields, we propose a practical approach to achieve skyrmion-bubble bundles with multi-$Q$ properties. This method is applicable to a wide range of traditional uniaxial magnets at room temperature[21, 24-27, 29]. We demonstrate the viability of topological transformations of skyrmion-bubble bundles using the wobbler and $B$-increasing techniques, which allow for increased and decreased $Q$, respectively. Our results highlight the diversity of topological spin textures in centrosymmetric uniaxial magnets, with excellent thermal stability at room temperature. This discovery could promote topological spintronic device applications based on the freedom of topological charges $Q$. Furthermore, the hexaferrite family is significant because it hosts multiferroic properties[38, 39], suggesting the potential for electrically-controlled topological spin textures in these materials.

## 4. Methods

*Preparation, structure characterization, and macroscopic magnetic measurement of bulk X-type $Sr_2Fe_{30}O_{46}$ hexaferrite crystal*: The hexaferrite crystals of $Sr_2Co_xFe_{30-x}O_{46}$ ($x \approx 2$) were grown by the flux method[34]. The powders of $SrCO_3$, $Fe_3O_4$, $Fe_2O_3$, and $Na_2CO_3$ were mixed and grounded thoroughly. The mixtures of raw materials were put into a platinum crucible and heated to 1593 K at a rate of 50 K/h and kept at this temperature for 48 h, then they were gradually cooled to 1313 K at a rate of 0.5 K/h, and finally cooled to room temperature at 150 K/h. All the procedures were performed in the air atmosphere. The synthesized single crystals with 1-2 centimeters in size and shiny surfaces were got by dissolving the flux in the hot nitric acid for several days.

The crystalline quality and crystallographic orientations of the shiny surfaces of grown crystals were confirmed by the Rigaku-TTR3 X-ray diffractometer using high-intensity graphite monochromatized Cu-Kα radiation. High-resolution transmission electron microscopy images and electron diffraction patterns of the shiny surfaces of grown crystals were all acquired by TEM (F200X, Talos) operated at 200 kV. The magnetic measurements were performed on a piece of cube-shaped single crystal (~4.7 mg) in Quantum Design Physical Properties Measurement System (PPMS) with Vibrating Sample Magnetometer Measurement (VSM) within the temperature range of 300–800 K. The magnetic fields were applied perpendicular to and parallel to the *c*-axis of the crystal sample, respectively.

*Fabrication of SrCoFeO lamellas*: The 150-nm thick SrCoFeO nanostructured lamellas were fabricated from a bulk single crystal using a standard lift-out method, with a focused ion beam and scanning electron microscopy dual beam system (Helios Nanolab 600i, FEI)[20, 51].

*TEM measurements*: We used in-situ Fresnel imaging in Lorentz-TEM (Talos F200X, FEI) with an acceleration voltage of 200 kV to investigate magnetic domains in the SrCoFeO lamellas. The TEM holder (model 636.6, Gatan) can support the varying temperature measurements.

*Micromagnetic simulations*: The zero-temperature micromagnetic simulations for magnetic domains in 150-nm thick lamellas were performed using MuMax3.[56] We consider in the Hamiltonian exchange interaction ($A$) energy, uniaxial magnetic anisotropy ($K_u$) energy, Zeeman energy, and dipole-dipole interaction energy.[56] Simulated magnetic parameters are set based on the material $Sr_2Co_2Fe_{28}O_{46}$ measured at 350 K ($K_u = 95.4$ kJ·m$^{-3}$, saturated magnetization $M_s = 366.6$ kA·m$^{-1}$) or at 300 K ($K_u = 81.5$ kJ·m$^{-3}$, saturated magnetization $M_s = 403.1$ kA·m$^{-1}$) . We set exchange constant $A = 8.25$ pJ·m$^{-1}$. The cell size was set at 3 × 3 × 3 nm$^3$.

**Supporting Information**

Supporting Information is available from the Wiley Online Library or from the author.

**Acknowledgments**

This work was supported by the National Key R&D Program of China, Grant No. 2022YFA1403603; the Natural Science Foundation of China, Grants No. 12174396, 12104123, 1197402, and 12241406; the National Natural Science Funds for Distinguished Young Scholar, Grants No. 52325105; Natural Science Project of Colleges and Universities in Anhui Province, Grant No. 2022AH030011; the Strategic Priority Research Program of Chinese Academy of Sciences, Grant No. XDB33030100; CAS Project for Young Scientists in Basic Research, Grant No. YSBR-084; the Research Fund of Hefei Normal University, Grant No. HXXM2022122; and the Equipment Development Project of Chinese Academy of Sciences, Grant No. YJKYYQ20180012.

**Author Contributions**

J.T., Y.W., and H.D. supervised the project and conceived the experiments. Y.W. synthesized the Sr-Co-Fe-O bulk single crystals and measured their crystals and macroscopic magnetism. J.T. performed the microdevice fabrication and TEM with the help of Y.W. and J.J.. J.T. performed the simulations. J.T., Y.W., and H.D. wrote the manuscript with input from all authors. All authors discussed the results and contributed to the manuscript.

**Conflict of Interest:**

The authors declare no competing financial interest.

Received: ((will be filled in by the editorial staff))
Revised: ((will be filled in by the editorial staff))
Published online: ((will be filled in by the editorial staff))

# Supporting Information:

## Skyrmion-Bubble Bundles in an X-type Sr$_2$Co$_2$Fe$_{28}$O$_{46}$ Hexaferrite above Room Temperature

*Jin Tang, Yaodong Wu\*, Jialiang Jiang, Lingyao Kong, Shouguo Wang, Mingliang Tian, and Haifeng Du\**

Jin Tang, Lingyao Kong, Mingliang Tian

School of Physics and Optoelectronics Engineering, Anhui University, Hefei, 230601, China

Yaodong Wu

School of Physics and Materials Engineering, Hefei Normal University, Hefei, 230601, China

E-mail: wuyaodong@hfnu.edu.cn

Jin Tang, Jialiang Jiang, Mingliang Tian, Haifeng Du

Anhui Province Key Laboratory of Condensed Matter Physics at Extreme Conditions, High Magnetic Field Laboratory, HFIPS, Anhui, Chinese Academy of Sciences, Hefei, 230031, China

E-mail: duhf@hmfl.ac.cn

Shouguo Wang

School of Materials Science and Engineering, Anhui University, Hefei 230601, China

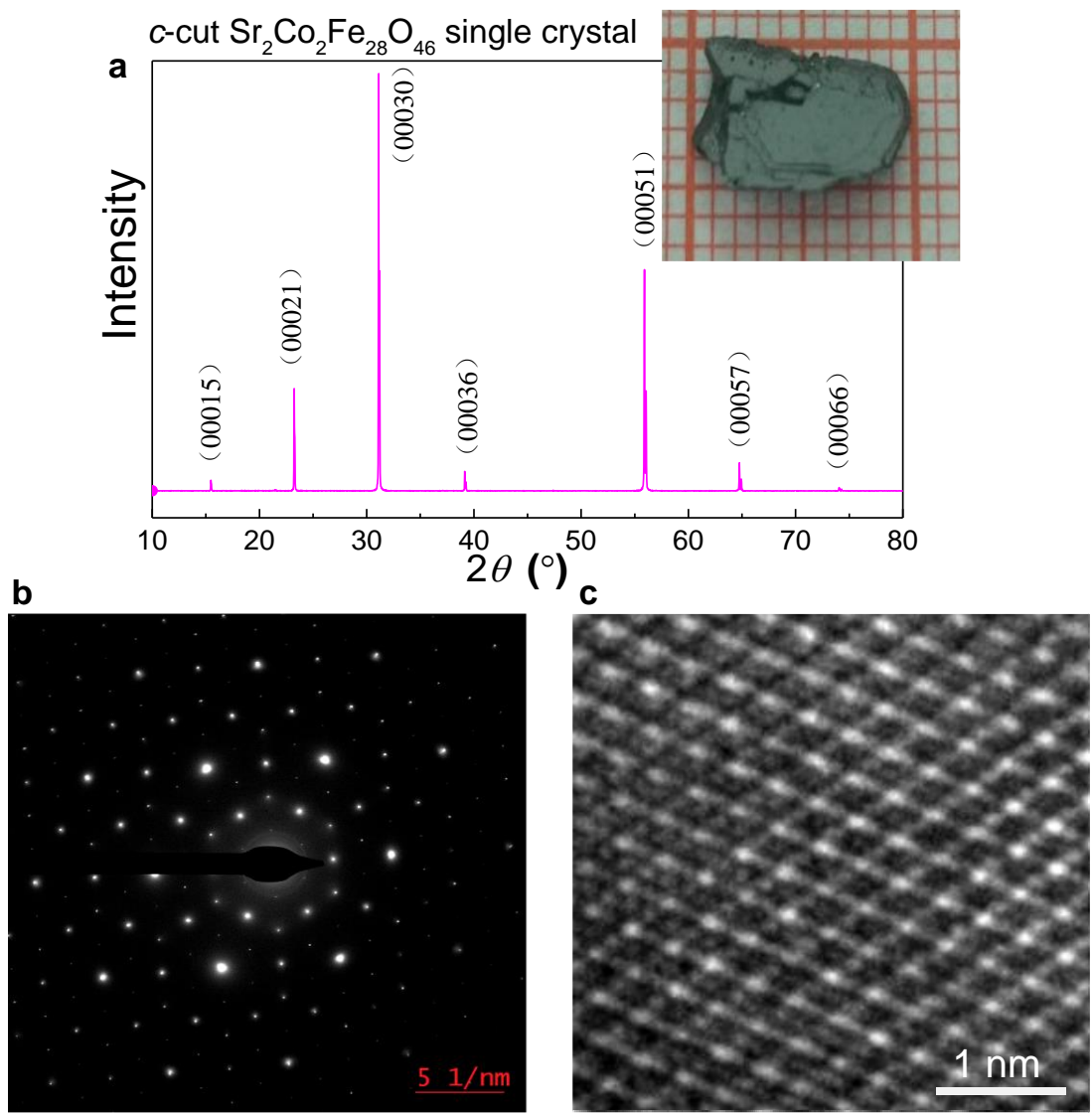

**Supplemental Figure S1. a)** X-ray diffraction of a *c*-cut SrCoFeO single crystal. Inset shows the bulk crystal. **b)** Electron diffraction of a *c*-cut SrCoFeO lamella. **c)** High-resolution imaging of a *c*-cut SrCoFeO lamella.

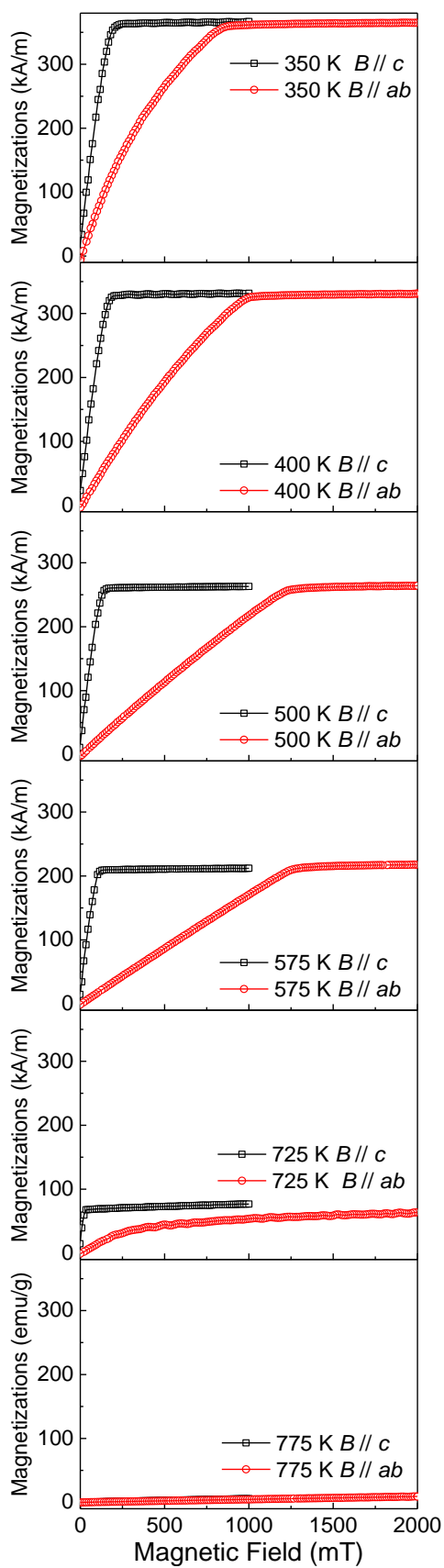

**Supplemental Figure S2.** Magnetization along the *c* axis and *ab* plane above room temperature.

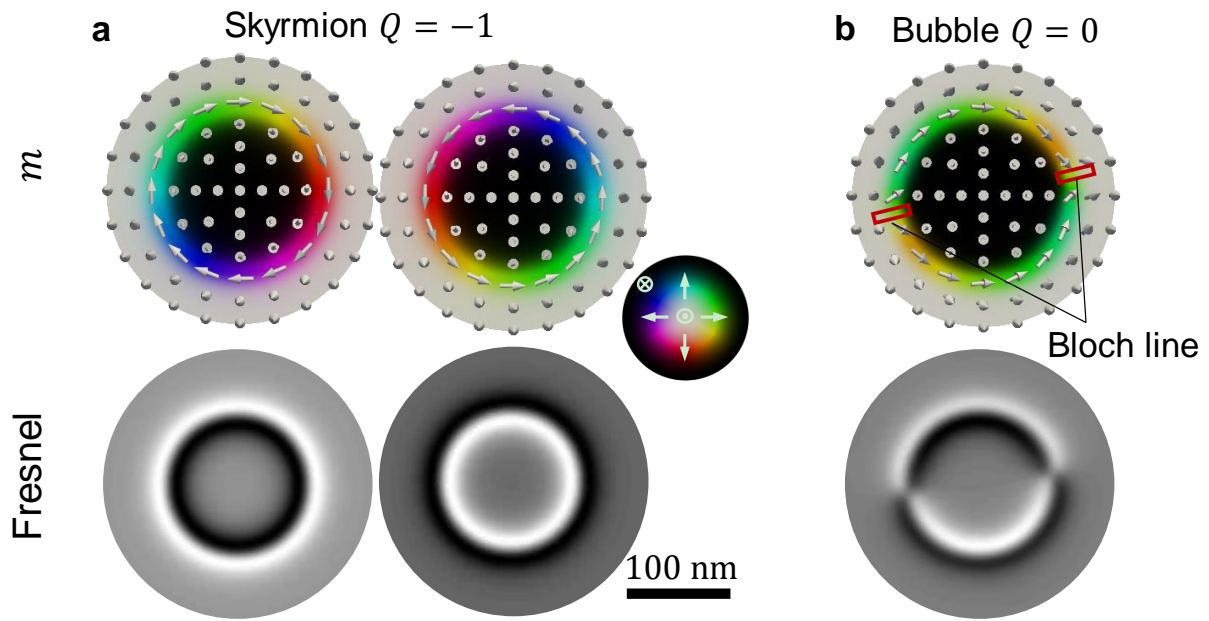

**Supplemental Figure S3.** Simulated spin configurations and corresponding defocused Fresnel contrasts of dipolar skyrmions a) with $Q = -1$ and bubble b) with $Q = 0$. The red wireframes in (b) indicate the locations of Bloch lines. The color represents the magnetization according to the colorwheel.

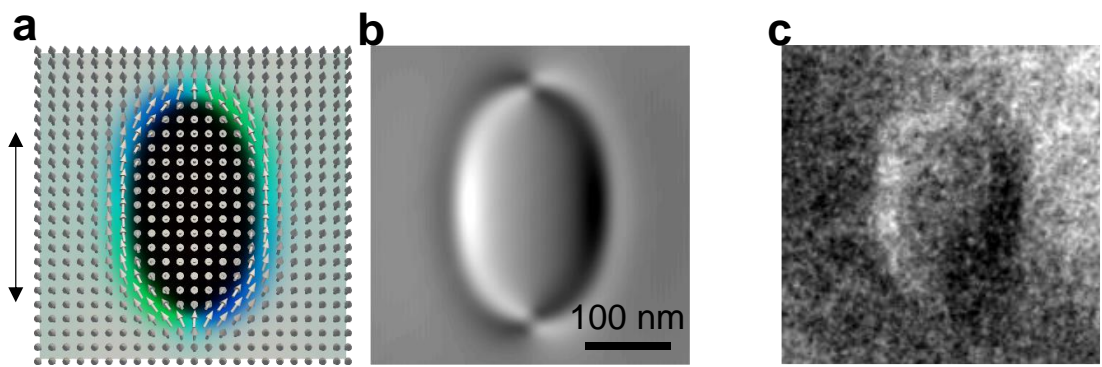

**Supplemental Figure S4.** Simulated magnetic configuration a) and corresponding under-defocused Fresnel image b) of a bubble in the presence of a tilted uniaxial anisotropy at $B = 225$ mT. The uniaxial anisotropy is $14.5°$ away from the out-of-plane direction. The in-plane orientation of uniaxial anisotropy is denoted by the arrow in (a). (c) Experimental under-defocused Fresnel image of a bubble at 200 K and at $B = 250$ mT.

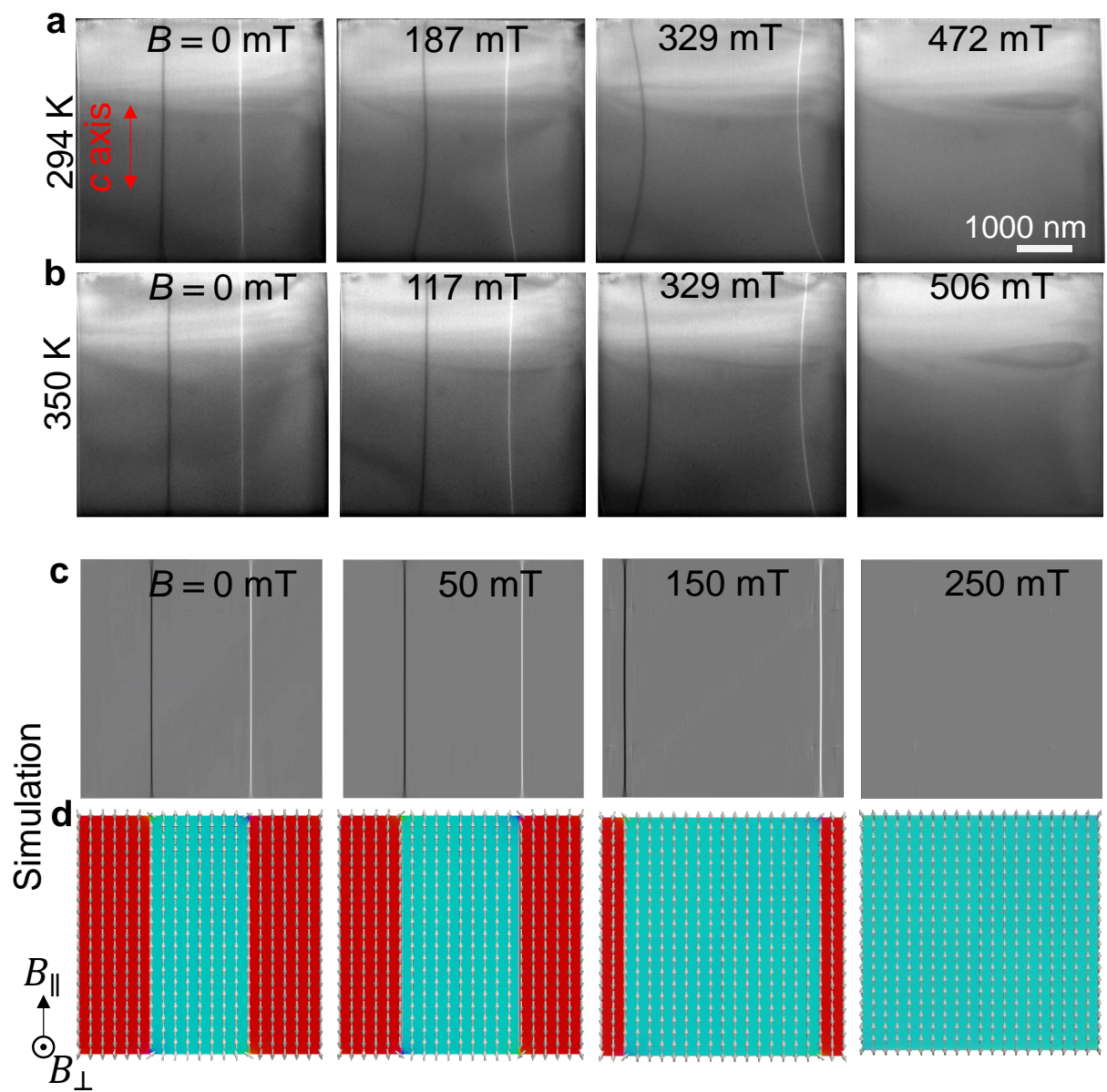

**Supplemental Figure S5.** Magnetic domains in a 150-nm thick lamella where the *c*-axis is parallel to the plane. a) Magnetic field-driven domain evolution at 294 K. b) Magnetic field-driven domain evolution at 350 K. Simulated under-defocused Fresnel images c) and spin configurations d) during the magnetic field-driven domain evolution. The orientation of the *c*-axis is denoted by a red arrow in (a). Noted the magnetic field is applied almost out-of-plane but with a slight 2.86° tilt toward the in-plane. The in-plane field component is denoted by the arrow in (d). When the field is applied, the region whose magnetization is parallel to the in-plane field component $B_{/\!/}$ expands, resulting in the displacement of domain walls.

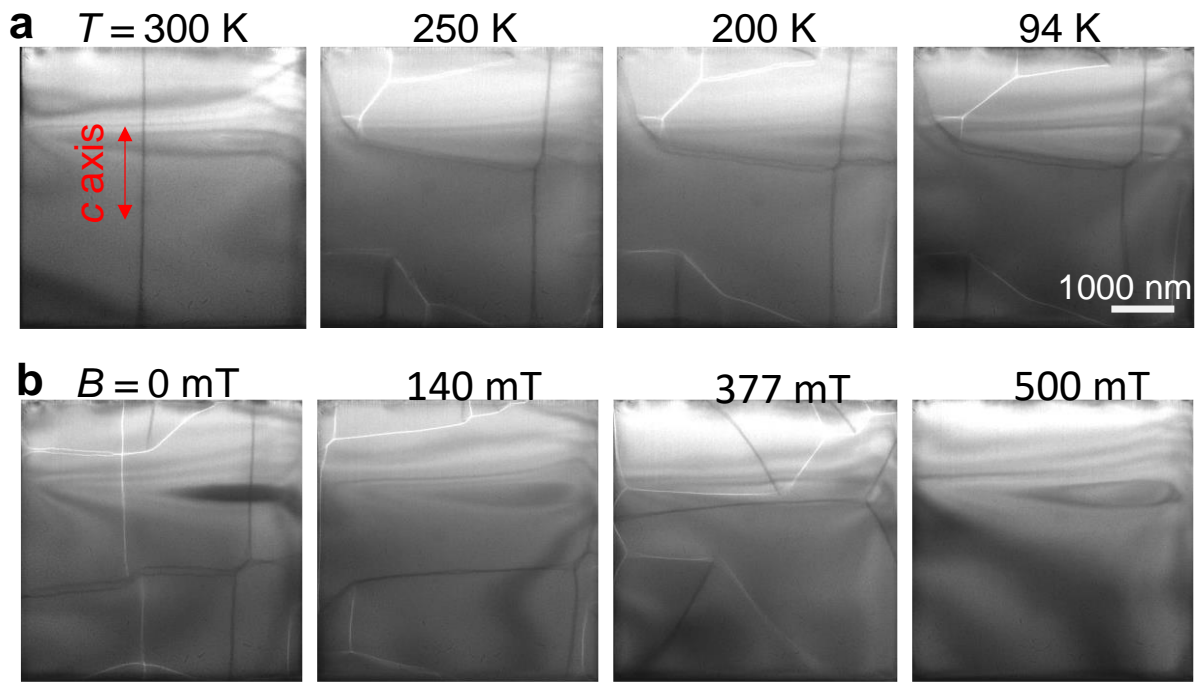

**Supplemental Figure S6.** Magnetic domains in a 150-nm thick lamella where the *c*-axis is parallel to the plane. a) Temperature-driven domain evolution at $B = 0$ mT. b) Magnetic field-driven domain evolution at 94 K.

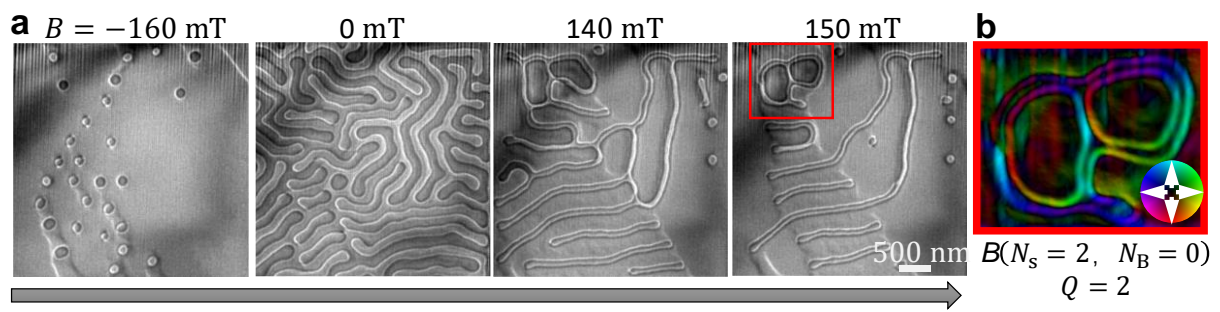

**Supplemental Figure S7.** a) Formation of Skyrmion-bubble bundles through the reversed fields. b) Retrieved in-plane magnetization mapping of a skyrmion-bubble bundle in the marked region of (a). Temperature $T = 294$ K. The color represents the in-plane magnetization according to the colorwheel.

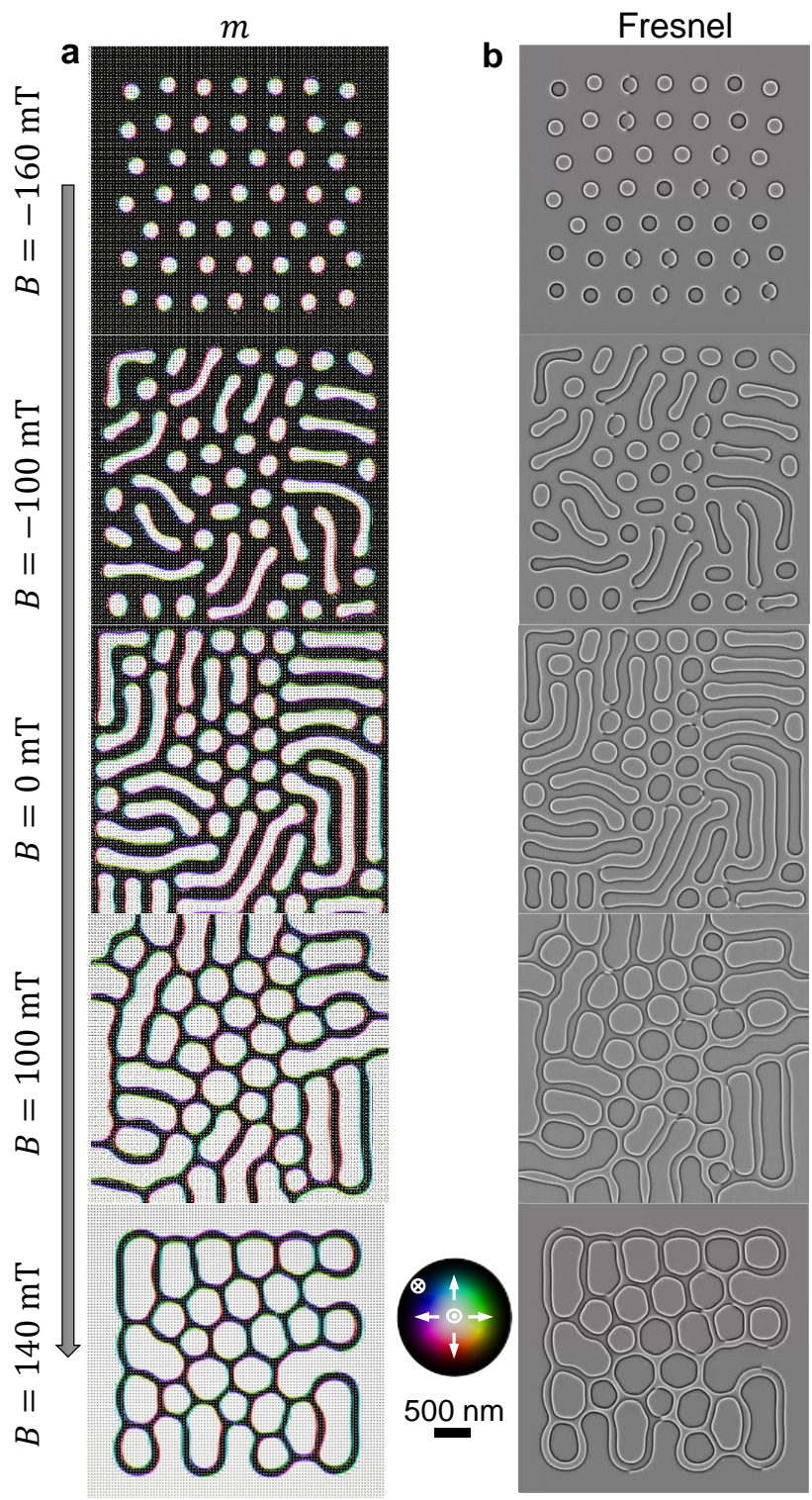

**Supplemental Figure S8.** Simulated magnetic evolution from skyrmion bubbles at negative fields to skyrmion-bubble bundles at positive fields. a) Simulated spin configurations. b) Corresponding simulated defocused Fresnel contrasts. The color represents the magnetization according to the colorwheel.

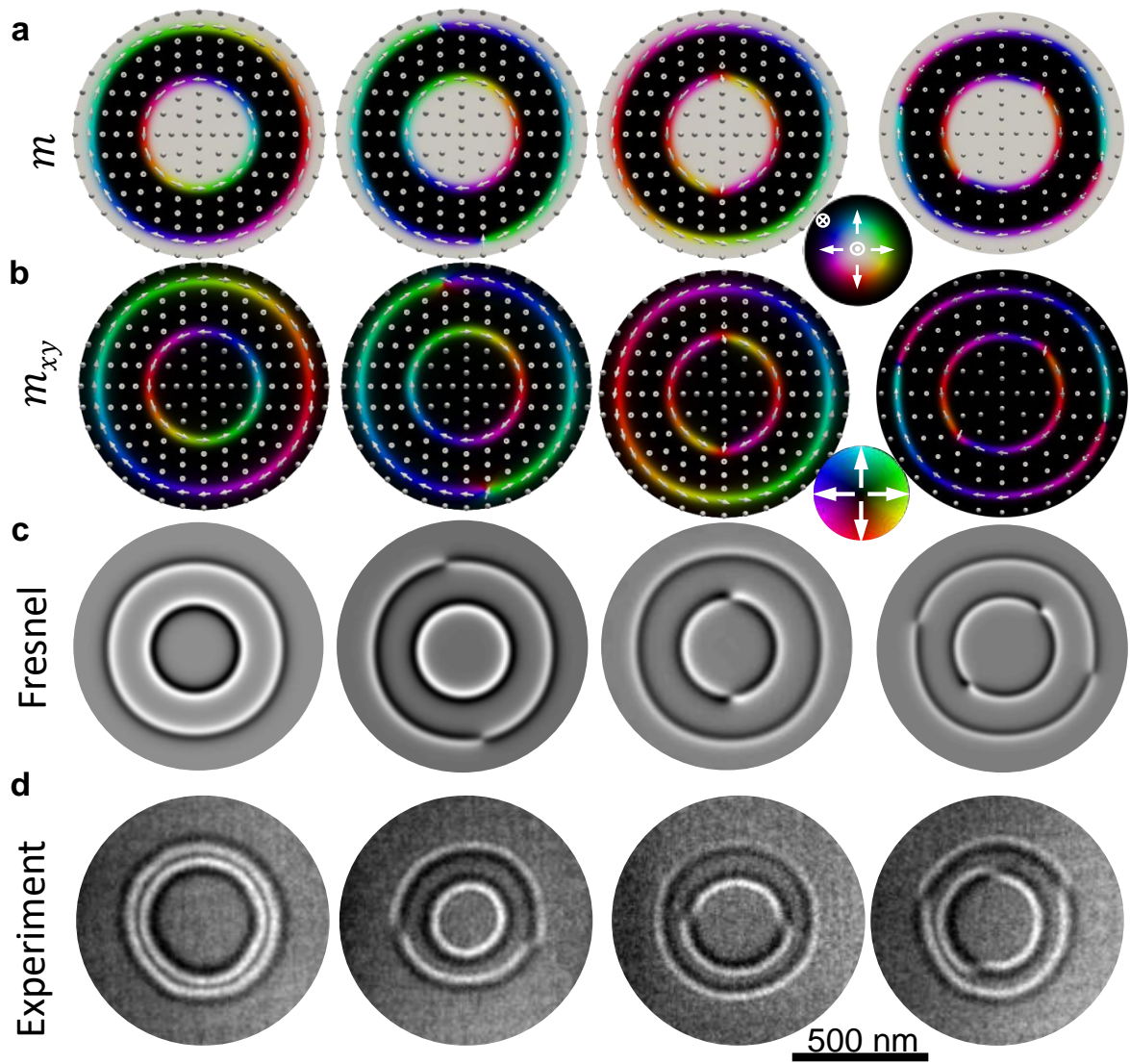

**Supplemental Figure S9.** Simulated magnetization a), in-plane magnetization b), and Fresnel images c) of 4 types of $2\pi$-vortices. d) corresponding experimental Fresnel images c) of 4 types of $2\pi$-vortices. Temperature $T = 350$ K. The color in (a) represents the magnetization based on the colorwheel in (a). The color in (b) represents the in-plane magnetization based on the colorwheel in (b).

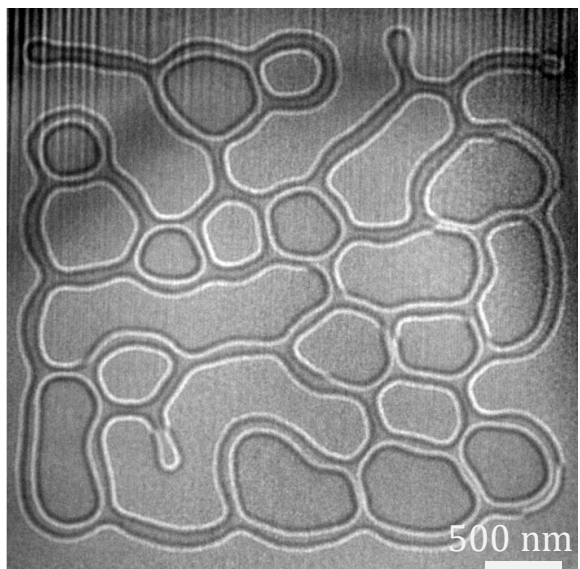

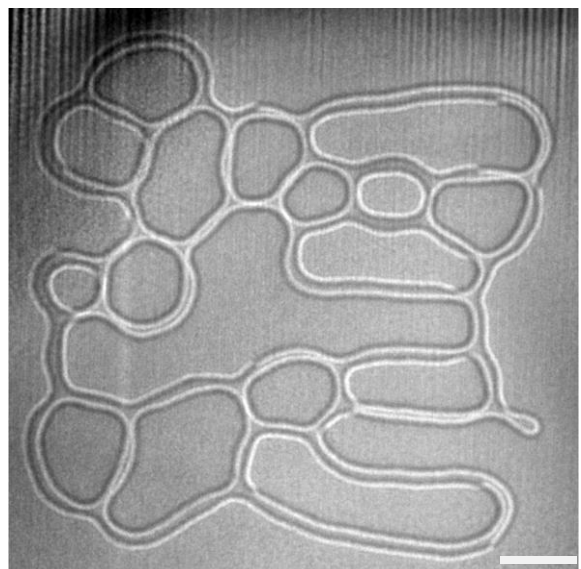

$B(N_\mathrm{s} = 14,\ N_\mathrm{B} = 7)$
$Q = 14$

$B(N_\mathrm{s} = 10,\ N_\mathrm{B} = 7)$
$Q = 10$

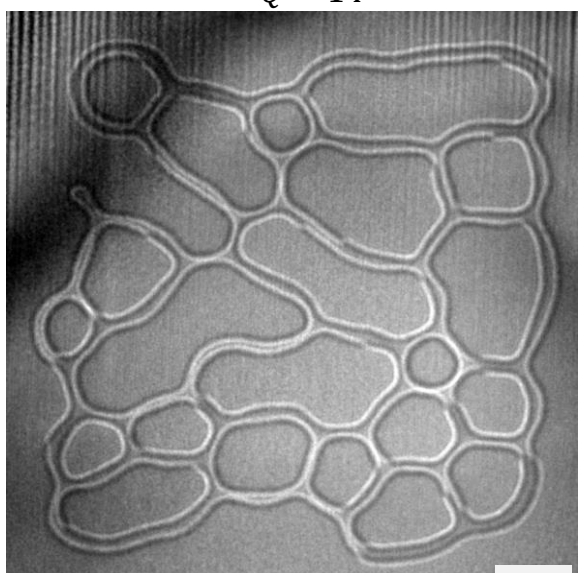

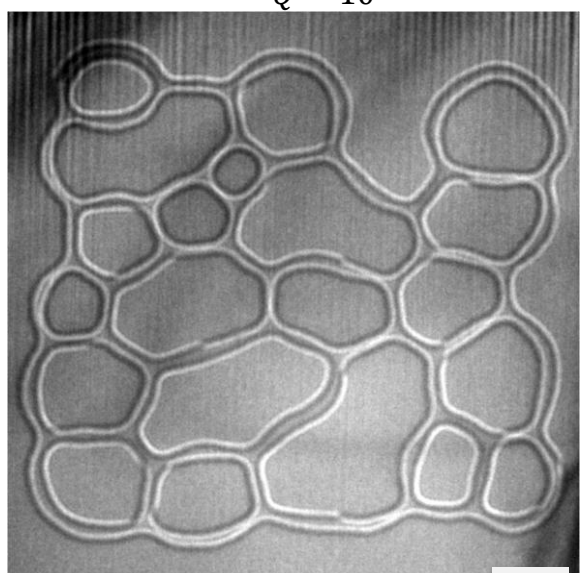

$B(N_\mathrm{s} = 5,\ N_\mathrm{B} = 17)$
$Q = 5$

$B(N_\mathrm{s} = 9,\ N_\mathrm{B} = 12)$
$Q = 9$

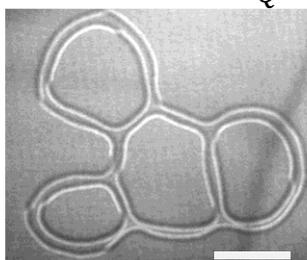

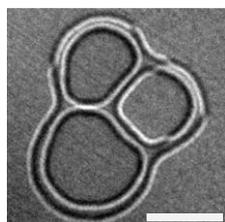

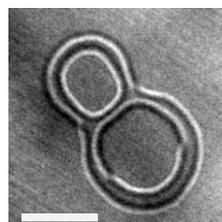

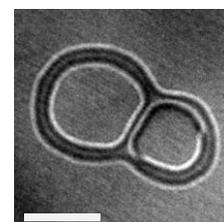

$B(N_\mathrm{s} = 2,\ N_\mathrm{B} = 1)$
$Q = 2$

$B(N_\mathrm{s} = 2,\ \overline{N_\mathrm{B} = 1})$
$Q = 2$

$B(N_\mathrm{s} = 1,\ N_\mathrm{B} = 1)$
$Q = 1$

$S(N_\mathrm{s} = 1,\ N_\mathrm{B} = 1)$
$Q = 0$

**Supplemental Figure S10.** Defocused Fresnel images of skyrmion-bubble bundles with a diversity of topological charges $Q$ and configurations. Temperature $T = 350$ K. Scale bar, 500 nm.

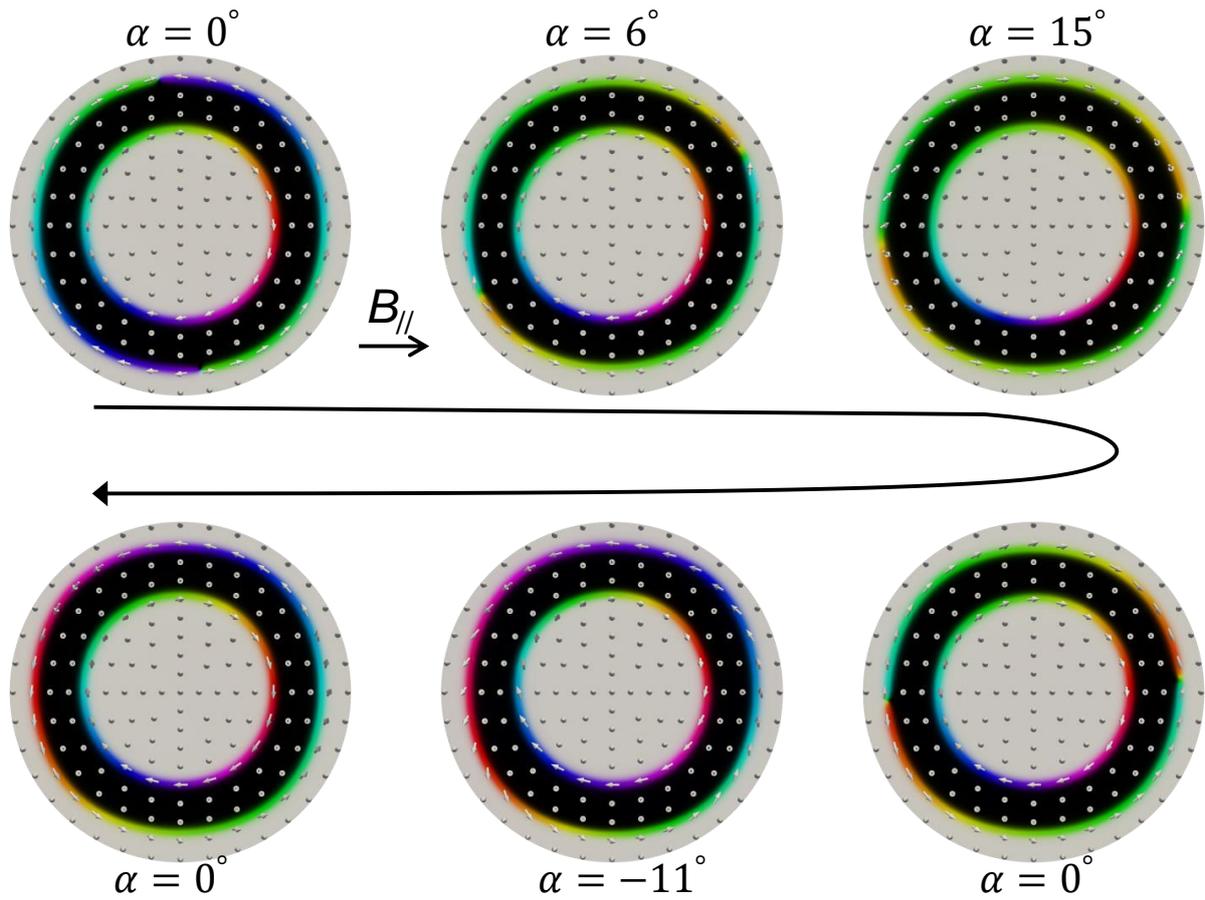

**Supplemental Figure S11.** Simulated topological magnetic transformation from a $B(N_S = 1,$ $N_B = 0)$ skyrmion-bubble bundle to a $S(N_S = 1, N_B = 0)$ skyrmion bundle through the wobbler operation. The arrow denotes the orientation of in-plane fields. Bloch lines move and could annihilate when modulating in-plane magnetic fields.

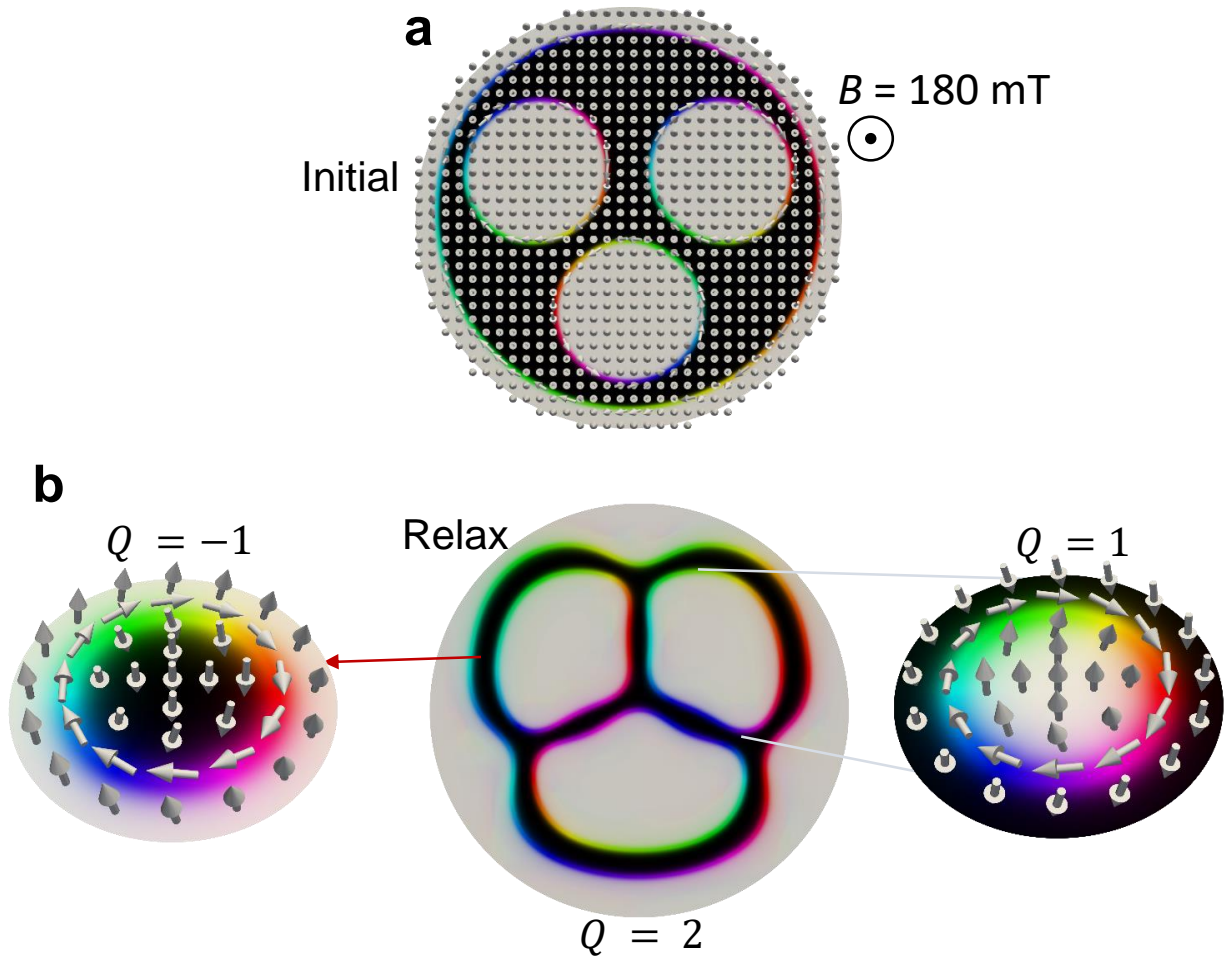

**Supplemental Figure S12.** a) Initial $Q = 2$ dipolar skyrmion bundle with internal loosely contacted skyrmions. The magnetic field is applied opposite to magnetizations between the internal skyrmions and outer boundary. b) Simulated equilibrium spin configuration of a $Q = 2$ dipolar skyrmion bundle relaxed from the initial non-equilibrium state in (a) to minimize Zeeman energy. The outer closed stripe domain is topologically equivariant to a skyrmion with $Q = -1$, whereas the internal 3 skyrmions contribute $Q = 3$.

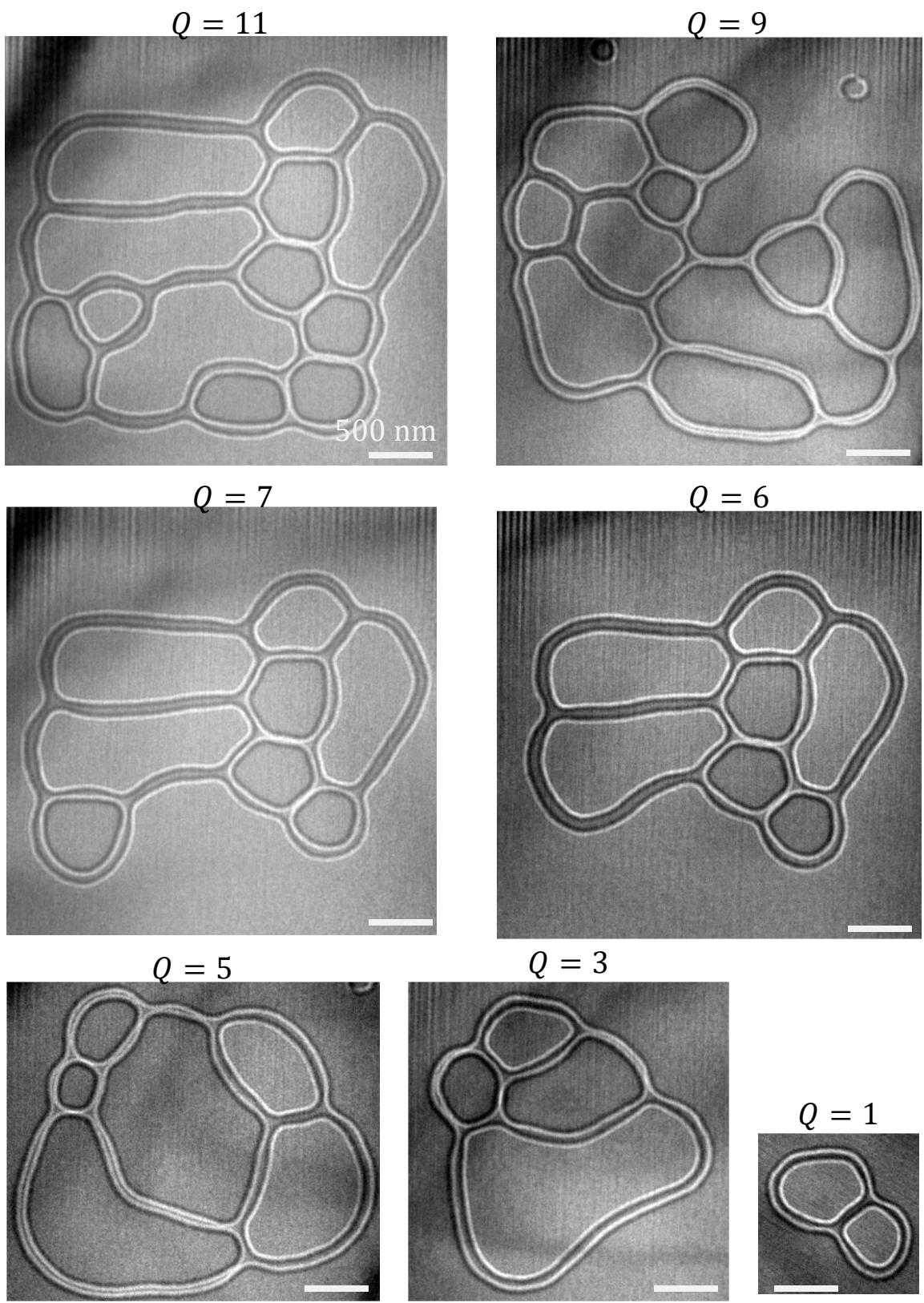

**Supplemental Figure S13.** Defocused Fresnel images of dipolar skyrmion bundles with a diversity of topological charges $Q$. Defocused distance, $-1000$ nm. Temperature $T = 325$ K. Scale bar, 500 nm.

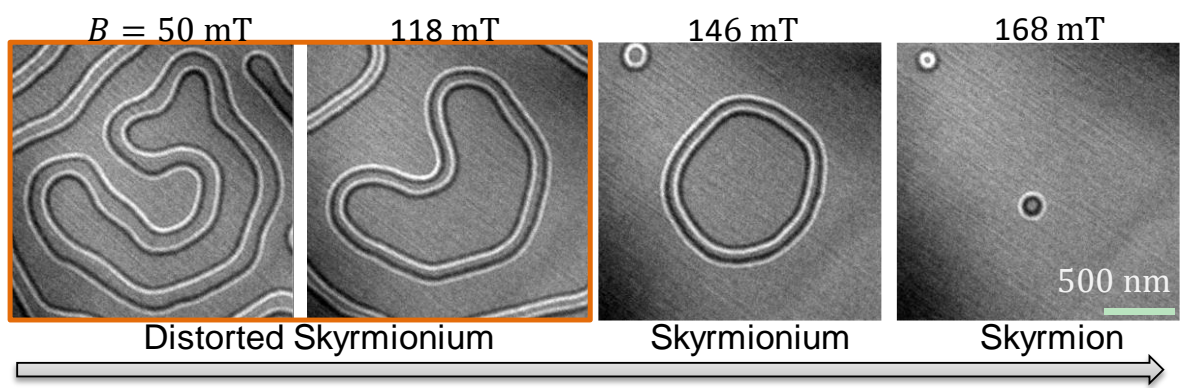

**Supplemental Figure S14.** Magnetic evolution from a distorted skyrmionium in the field-increasing process at temperature $T = 360$ K.